\documentclass{aa}

\usepackage{graphicx}
\usepackage{txfonts}
\usepackage[utf8]{inputenc}
\usepackage{color}
\def\gr{$\gamma$-ray}

\begin{document}

\title{Measurement of size of gamma-ray source in blazar B0218+357 from microlensing at 100 GeV energy}

\author{Ie.Vovk\inst{1} \and A. Neronov\inst{2}}
\institute{
  Institute for Cosmic Ray Research, The University of Tokyo, 5-1-5 Kashiwa-no-Ha, Kashiwa City, Chiba, 277-8582, Japan \\
  \and
  ISDC, Astronomy Department, University of Geneva, Ch. d'Ecogia 16, 1290, Versoix, Switzerland
}

\authorrunning{Ie.Vovk \& A. Neronov}
\titlerunning{Microlensing of B0218+357 in VHE band}
   
%


 
\abstract
{Observations of the effect of microlensing in gravitationally lensed quasars can be used to study the structure of active galactic nuclei on distance scales down to the sizes of the supermassive black holes powering source activity.}
{We search for the microlensing in the signal from a gravitationally lensed blazar B0218+357 in very-high-energy \gr\ band.}
{We combine observations of a bright flare of the source in 2014 with Fermi/LAT and MAGIC telescopes in 0.1-100~GeV energy range. Using the time-delayed leading and trailing signals from two gravitationally lensed images of the source, we measure magnification factor at the moment of the flare. We use the scaling of the maximal magnification factor with the source size to constrain the size of \gr\ emission region in the entire 0.1-100~GeV energy range.}
{The magnification factor in the very-high-energy band derived from a comparison of Fermi/LAT and MAGIC data is $\mu_{VHE} = 36^{+40}_{-26}$, substantially larger than that in the radio band. This suggests that one of the source images is strongly affected by microlensing at the moment of the flare. Assuming that the microlensing is produced by a stellar mass object in the lens galaxy, we constrain the size of the emission region in the $E>100$~GeV band to be $\mathrm{R_{VHE} = 2.2^{+27}_{-1.7} \times 10^{13}~cm}$. We note that the spectrum of the microlensed source was unusually hard at the moment of the flare and speculate that this hardening may be due to the energy dependent microlensing effect. This interpretation suggests that the source size decreases with energy in 0.1-100~GeV energy range studied.}
{}


\maketitle


\section{Introduction}

Very-high-energy (VHE) emission from Active Galactic Nuclei (AGN) can be affected by the effect of pair production on low-energy photon backgrounds \citep{PhysRev.155.1404} generated by accretion flows onto  supermassive black holes. Conventional models of accretion on supermassive black holes at rates close to the Eddington limit suggests that emission from optically thick accretion flow at the distance $\mathrm{R}$ from the black hole has temperature $\mathrm{k_BT\sim 10[M/_{bh}10^8M_\odot]^{1/4}[R/R_g]^{-1/2}~eV}$~\citep{1973A&A....24..337S,Neronov:2019uht}, where $\mathrm{M_{bh}}$ is the black hole mass and $\mathrm{R_g\simeq 1.5\times 10^{13}[M_{bh}/10^8M_\odot]~cm}$ is the gravitational radius. Dense $\mathrm{\epsilon\sim 10~eV}$ photon background around the black hole can block the escape of \gr s with energies $\mathrm{E_\gamma\sim 100[\epsilon/10~eV]^{-1}~GeV}$ because such \gr s produce electron-positron pairs in interactions with low-energy photons.  Scattering of the $\epsilon\sim 10$~eV  photons in the Broad Line Region (BLR) \citep{blr} may affect propagation of \gr s on the scales $\mathrm{R_{BLR}\sim 10^{17}~cm~\gg R_g}$ and at energies down to GeV band \citep{2010ApJ...717L.118P}. Model calculation of propagation of VHE \gr s through  radiation fields created by the accretion flow and scattered in the BLR lead to a conjecture that \gr s can only escape from a region outside the BLR situated in the parsec-scale jet ejected by the black hole \citep{2009MNRAS.397..985G}.

This conjecture is challenged by observations of fast variability of VHE \gr\ signals from some AGNs~ \citep{Aharonian:2007ig,Berge:2006uls,2014Sci...346.1080A,2013ApJ...767..103V}. Causality argument suggests that the size of electron acceleration region is  limited to be $\mathrm{R<ct_{var}\sim 10^{14}[t_{var}/1~hr]~cm}$, if measured in the reference frame of the black hole \citep{Neronov:2008ih}.  Attempts to reconcile the short variability time scales with pair production constraints on the emission region size include models with extremely compact emission regions embedded into extended parsec-scale jets \citep{2009MNRAS.395L..29G}. Origin of such very compact regions remains obscure. 

VHE \gr s can escape from the vicinity of the black hole if the accretion proceeds through a Radiatively Inefficient Accretion Flow (RIAF) with luminosity orders of magnitude below the Eddington limit \citep{1982Natur.295...17R,1994ApJ...428L..13N}. Such accretion flows are found in nearby radio galaxies and further away BL Lac type objects. RIAF luminosity is typically generated by optically thin synchrotron emission from mildly relativistic electrons in the accretion flow \citep{2014ARA&A..52..529Y}. This emission is in the infrared band and it does not present an obstacle for the escape of VHE \gr s. In this cases fast-variability of \gr\ emission can be readily explained by particle acceleration close to the black hole at the base of the jet \citep{2000PhRvL..85..912L,2003NewAR..47..693N,2007ApJ...671...85N,2014Sci...346.1080A,2016A&A...593A...8P}.

It is difficult to determine the accretion mode on source-by-source basis in blazars, because both the infrared emisssion from RIAF and optical-to-ultraviolet emission from optically thick accretion flow can be hidden behind much stronger Doppler-boosted emission from the relativistic jet. Only some brightest blazars of Flat Spectrum Radio Quasar (FSRQ) type reveal characteristic ``big blue bumps'' attributed to the optically thick accretion disks \citep{1998A&ARv...9....1C,2004ASPC..311...37W}. Thus, it is not possible to directly measure the optical depth of the accretion flow and in this way constrain the location of the VHE \gr\ emission region by measuring the distance at which the optical depth for the pair production decreases below unity.  

Nevertheless, a direct measurement of the size of emission region at different wavelengths is possible in a special case of gravitationally lensed AGN, through the effect of microlensing \citep{2011ApJ...729...34B,2010ApJ...709..278D}. Passage of stars in the lensing galaxy through the line of sight toward one of the images of the gravitationally lensed source leads to temporary magnification of the image with the magnification factor that depends on the source size.

Two gravitationally lensed blazars are detected in \gr s: PKS 1830-211 \citep{2015ApJ...799..143A} and QSO B0218+357 \citep{2014ApJ...782L..14C}. In both sources microlesing has been observed  and proved to be efficient in constraining the size of the high-energy \gr\ source in 0.1-10 GeV energy range \citep{2015NatPh..11..664N, b0218_vovk_neronov}. 

Here we report detection of the microlensing effect in  B0218+357 in the VHE band, at the energy $E_\gamma\sim 100$~GeV. This source at redshift $z=0.944\pm 0.002$ \citep{2000ApJ...545..578C} has two images, A and B \citep{1995MNRAS.274L...5P} with the flux from the image B delayed by $11.4\pm 0.2$~d with respect to the image A \citep{2018MNRAS.476.5393B}. This allows to distinguish contributions of the images A and B to the \gr\ flux even if the two images are not resolved in the \gr\ band. We derive a measurement of the magnification factor from a combination of observations of a bright flare of two images of the source with space-based Fermi/LAT~\citep{Fermi_LAT,2014ApJ...782L..14C} and ground-based MAGIC \gr\ telescopes~\citep{2016A&A...595A..98A}. Comparing the VHE \gr\ magnification factor measurement with that in the radio band (not affected by the microlensing) we are able to detect the microlensing effect and constrain the size of the \gr\ source.

\section{Data analysis}

We use the publicly available  Pass~8 reprocessed photon data set from the Fermi/LAT telescope\footnote{http://fermi.gsfc.nasa.gov/cgi-bin/ssc/LAT/LATDataQuery.cgi}. The data were analysed using the Fermitools package and FermiPy framework\footnote{\url{https://fermi.gsfc.nasa.gov/ssc/data/analysis/}} v0.17.3~\citep{fermipy} as described in FermiPy documentation\footnote{\url{https://fermipy.readthedocs.io/}}, using the {\tt SOURCE} event class and {\tt P8R3 P8R3\_SOURCE\_V3} response functions. The photons were selected from the $20^\circ$ region around the position of B0218+357. The fluxes of all sources in the selected region were estimated from the likelihood fit, which included all the sources from the 4FGL~\citep{4FGL} catalogue within 35 degrees from B0218+357 as well as the galactic ({\tt gll\_iem\_v07.fits}) and extragalactic ({\tt iso\_P8R3\_SOURCE\_V3\_v1.txt}) diffuse backgrounds.

We focus on a flaring episode of the source in the MJD~56851--56865 range, during which a hard source spectrum was reported in the high-energy band 0.1-10~GeV~\citep{B0218+35_Fermi_paper_new}. Only the leading flare from the image A (MJD~56851.62--56853.62) was significantly detected by Fermi/LAT during this episode.  The trailing flare (from the image B) has not been detected by Fermi/LAT.  The measured flux above 100~MeV during the expected trailing flare B (MJD~56863.08--56865.08) remained consistent with the non-flaring state~\citep{b0218_vovk_neronov}.  The trailing flare B was previously detected by MAGIC telescope in the energy range above 100~GeV \citep{2016A&A...595A..98A}. However, MAGIC observations did not cover the leading flare A and hence a direct measurement of the magnification factor in the VHE band was not possible. 

In the GeV band, a non-negligible contribution from the leading flare may be expected during the expected time window of flare B. In order to account for it when searching for the flare B emission with Fermi/LAT, we have combined in a joint fit the time intervals including the expected trailing flare (MJD~56863.08--56865.08), in-between the flares (MJD~56853.62--56863.08) and after the trailing flare (MJD~56865.08--56875.08). During the fit, we have allowed for an additional log-parabola component during the first time interval to account for the putative flare B contribution, while the remaining sources, including the baseline model for the flare A, were jointly optimized over all three intervals.

\begin{figure}
    \includegraphics[width=\linewidth]{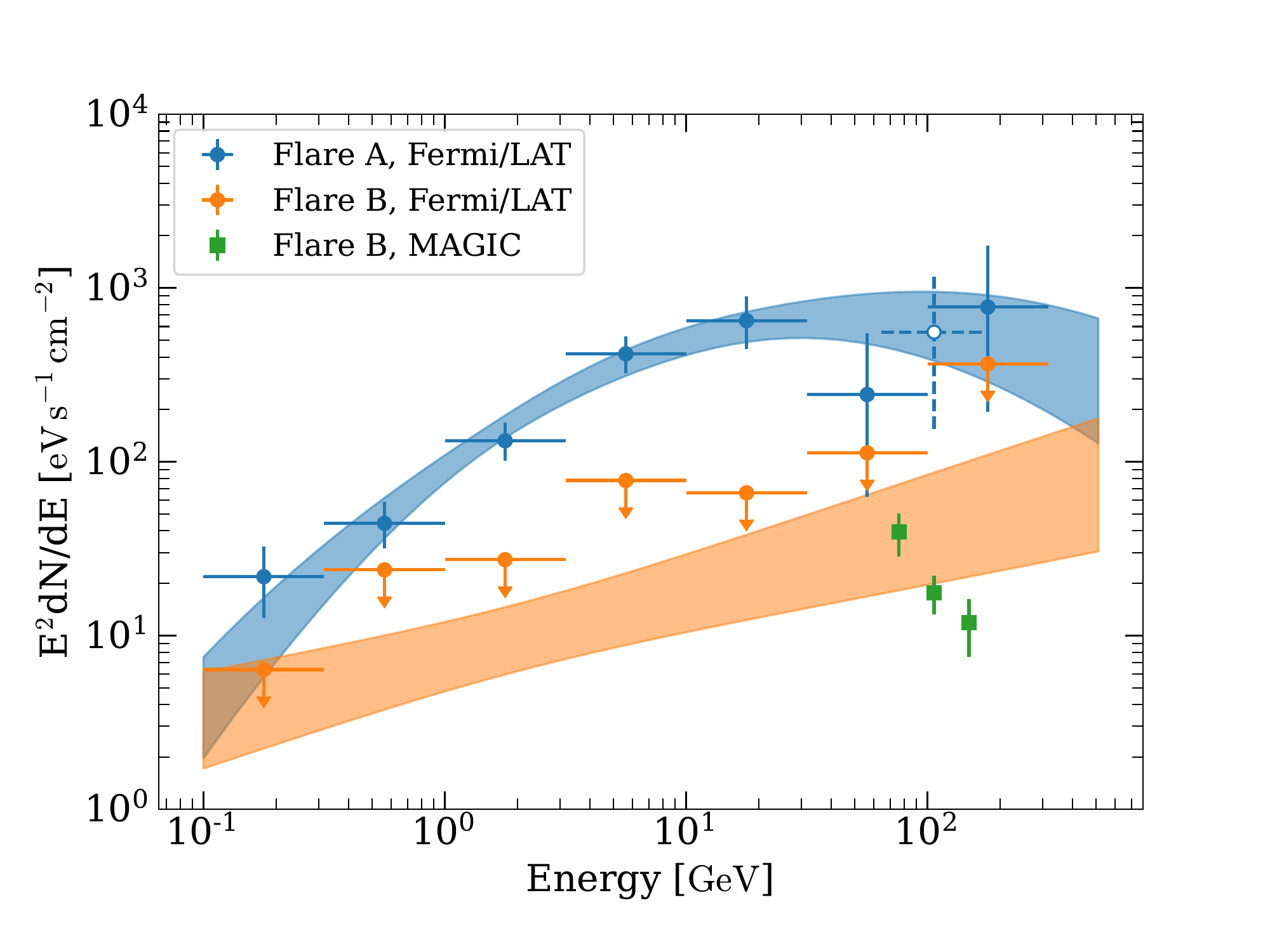}
    \caption{
        Reconstructed spectral energy distribution of B0218+357 during the MJD~56851--56865 flaring episode from the Fermi/LAT data. Blue data points and band represent the leading flare emission derived here, whereas the orange ones correspond to the trailing emission component. Green data points represent the MAGIC measurements of the trailing component~\citep{2016A&A...595A..98A}. An open circle data point represent the Fermi/LAT flux extracted in the $\mathrm{65-175~GeV}$ energy range of the MAGIC measurements.
    }
    \label{fig::sed}
\end{figure}

The result of our re-analysis of the broadband \gr\ spectrum of the source during the leading and trailing flares is shown in Fig.~\ref{fig::sed}. The derived spectrum of the leading flare extends up to 300~GeV ($TS \approx 25$ in the $100-300$~GeV energy bin) and is well described by a log-parabola shape
\begin{equation}
    \frac{dN}{dE} = N_0 \left(
        \frac{E}{E_b}
    \right)^{
        \alpha + \beta \log{(E/E_b)}
    }
    \label{eq::logparabola}
\end{equation}
with $\mathrm{N_0 = (1.14 \pm 0.27) \times 10^{10}~ph/cm^2~s~MeV}$, $\beta=-0.14 \pm 0.06$, the break energy fixed to the 4GLF value of $E_b=778$~MeV and a remarkably hard index of $\alpha=-0.88\pm0.26$. When fit with a power law spectral model, the slope of the leading flare in the $0.1-30$~GeV energy range is $\Gamma=-1.16 \pm 0.20$. The trailing flare B, at the same time, was indeed not detected on top of the baseline emission during the expected time window. The derived flux upper limits at 95\% confidence level are consistent with the MAGIC detection at the same energies~\citep{2016A&A...595A..98A}.

Detection of the VHE \gr\ flux from the leading flare with Fermi/LAT enables a straightforward measurement of the VHE band magnification factor ratio via comparison of the flare A flux with Fermi/LAT and flare B flux with MAGIC, as described below.


\section{Microlensing  in the VHE band}

The spectral energy distributions of the leading / trailing flares, depicted in Fig.~\ref{fig::sed}, enable gravitational lensing magnification factor $\mathrm{\mu}$ estimate in several ways. 

A straightforward estimate of magnification in the $\mathrm{0.1-100~GeV}$ energy band can be obtained from the ratio of leading to trailing components extracted fluxes in the separate energy bins. To properly account for the measurement uncertainties, we have used the log-likelihood profiles, automatically extracted by FermiPy as function of the flux normalization for every data point. Parametrizing the trailing and leading fluxes as $\mathrm{F}$ and $\mathrm{\mu F}$, we have used the total log-likelihood $\mathrm{L(F, \mu)}$, marginalized over $\mathrm{F}$, to estimate the lower limits on $\mathrm{\mu}$ at the 90\% confidence level. The energy-dependent magnification bound derived this way reaches $\mu \gtrsim 17$ in the $\mathrm{10-31.6~GeV}$ energy bin.

Combining the Fermi/LAT and MAGIC measurements, it is also possible to directly measure the magnification factor in the $\mathrm{65-175~GeV}$ energy range where VHE emission of flare B was detected. Here we compare a weighted average of the MAGIC data points with the Fermi/LAT measurement in that entire energy range to compensate for its lack of sensitivity in the individual MAGIC energy bins. Using the same procedure as above -- but this time using the gaussian likelihood for the combined MAGIC data point, -- we have found $\mu = 36^{+40}_{-26}$ in this energy range (uncertainties correspond to the 68\% confidence interval).

Finally, the remarkable hardness of the leading flare emission in the $\mathrm{0.1-30~GeV}$ energy range may itself be attributed to the gravitational lensing phenomenon, modifying the spectrum in the energy-dependent manner. \gr\ spectra of blazars, roughly represented by the power law $dN_\gamma/dE_\gamma\propto E^{-\Gamma}$, are characterised by spectral slopes $\Gamma>1.5$, with rare exceptions like the flaring state of Mrk~501 \citep{2012A&A...541A..31N}. Assuming the intrinsic source spectrum has a hardest conventionally assumed spectral index of $\Gamma_{min} = 1.5$, the energy-dependent gravitational lensing influence can be estimated from the ratio of the measured source fluxes to the $\mathrm{dN/dE \sim E^{-\Gamma_{min}}}$ approximation. It should be noted though, that such an estimate, based on the leading flare A emission only, reflects only the energy dependency of the magnification factor $\mu$, but not its overall scale. The latter needs to be derived separately; here we normalize the obtained scaling to the magnification factor in the $\mathrm{65-175~GeV}$ energy range, derived above.

The magnification factor ratio derived from comparison of Fermi/LAT and MAGIC data is larger than the radio band magnification factor ratio between images A and B, $\mu_r\simeq 4$ \citep{2007A&A...465..405M}, which can be attributed to the influence of microlensing~\citep{b0218_vovk_neronov}. The microlensing can hardly affect the radio source that is resolved and is known to have the multi-parsec size. Such a large source cannot be magnified  by the effect of microlensing on stars in the lensing galaxy that affects only sources with the sizes smaller than the size of the Einstein ring, 
\begin{equation}
    R_E=
        \sqrt{
            \frac{4GM}{c^2}\frac{D_{LS} D_S}{D_L}
        }
        \simeq
        3\times 10^{16} \left[
            \frac{M}{M_\odot}\right]^{1/2}\left[\frac{D}{1\mbox{ Gpc}}
        \right]^{1/2}\mbox{ cm}
\end{equation}
where $D_L,D_S,D_{LS}$ are distances to the lens, source, between the lens and the source, $D\sim D_L \sim D_{LS}$ is order-of-magnitude distance estimate and $M$ is the mass of the object responsible for the microlensing. 
The microlensing could have affected the \gr\ source flux during the flare if the \gr\ source size is smaller than $R_E$.

\begin{figure}
    \includegraphics[width=\linewidth]{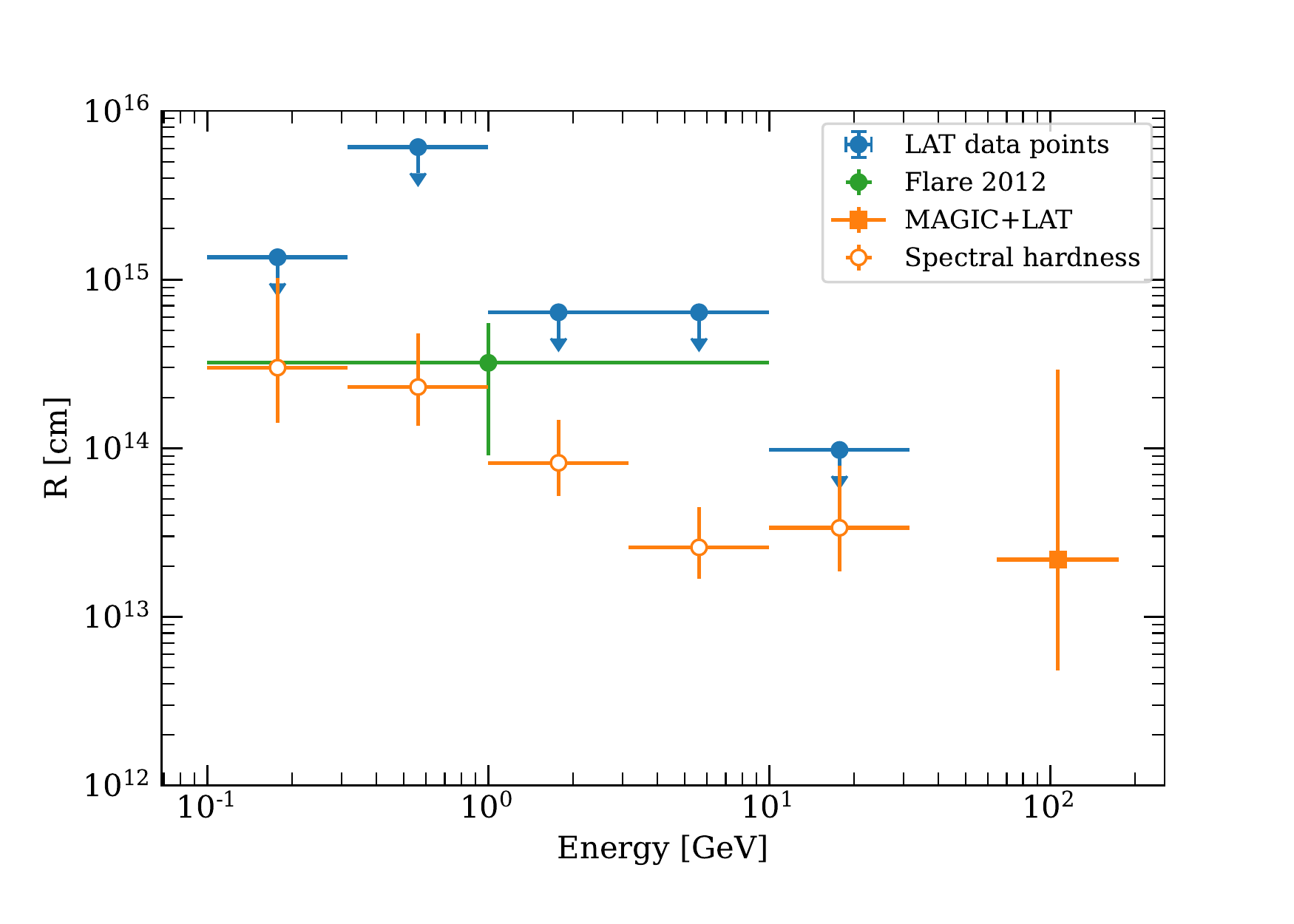}
    \caption{
        Reconstructed emission region size in B0218+357 as a function of energy. Blue upper limits were derived from the ratio of the leading and trailing components fluxes; solid orange data point represents the source size measurement stemming from Fermi/LAT and MAGIC detection of the $\mathrm{\sim 100~GeV}$ emission from the leading and trailing components correspondingly. Open orange circles represent the source size energy scaling, required to explain the extreme hardness of the measured leading component spectrum. Green data point marks the source size measurement during the 2012 flare of B0218+357~\citep{b0218_vovk_neronov}.
    }
    \label{fig::size}
\end{figure}

The relation between the microlensing magnification factor and source size is 
\begin{equation}
\label{eq:mu_ml}
    \mu_{ml}=\sqrt{\frac{R_E}{R_\gamma}}\simeq 10\left[\frac{R_\gamma}{3\times 10^{14}\mbox{ cm}}\right]^{-1/2}\left[\frac{M}{M_\odot}\right]^{1/4}\left[\frac{D}{1\mbox{ Gpc}}\right]^{-1/4}
\end{equation}
Comparing the radio magnification factor with the one from the \gr\ band during the flare one can find an estimate of the microlensing effect $\mu_{ml}=\mu_{vhe}/\mu_r$ and using Eq.~\ref{eq:mu_ml} estimate the source size in the $\mathrm{0.1-100~GeV}$ \gr\ band. The result of such estimate is shown in Fig. \ref{fig::size}. One can see that the size of the VHE emission region is found to be very compact with $\mathrm{R_{VHE} = 2.2^{+27}_{-1.7} \times 10^{13}~cm}$.

The source size estimate in the VHE band is somewhat smaller than in the high-energy (0.1-10~GeV) band. This indicates a possibility that source size is actually shrinking with increasing energy. Such decreasing source size may well be responsible for the unusually hard spectrum of the source observed during the leading flare. Indeed, as one can see from Fig.~\ref{fig::size}, the required source size scaling is consistent with the measurements in HE and VHE bands.

\section{Discussion}

Compactness of the VHE \gr\ source inferred from the observation of the microlensing effect may pose a problem of opacity of the \gr\ source to the pair production on low
energy photon backgrounds from the accretion flow.  In the case of  B0218+357 the parameters of the accretion flow are not well constrained. It is not clear at what rate does the source accrete because no spectral signatures of either radiatively efficient accretion at near-Eddington rate (the "big blue bump" in the optical / ultraviolet spectrum) or of a RIAF (a bump due to synchrotron emission in the infrared spectrum) are detectable. Modelling  of \citet{2022MNRAS.510.2344A} suggests that infrared-to-ultraviolet flux from the AGN is dominated by the synchrotron emission from relativistic jet, supposedly from its part at distances from $R_{jet}\sim 2\times 10^{17}$~cm to $3\times 10^{20}$~cm from the black hole. 

\gr s arriving at Earth with energies $E_\gamma\simeq 100$~GeV are emitted from the source with energies $E'=E(1+z)\simeq 200$~GeV. Such photons are most efficiently absorbed in pair production on blue / ultraviolet photons with energies $\epsilon'\simeq 5$~eV that reach the Earth with energy $\epsilon=\epsilon'/(1+z)\simeq 2.5$~eV.  An estimate of the powerlaw flux at 2.5~eV is $F_{v}\simeq 2\times 10^{-14}$~erg/cm$^2$s \citep{2022MNRAS.510.2344A}, so that the intrinsic source luminosity is $L_{v}=4\pi D_L^2F_{v}\simeq 10^{44}$~erg/s. If this emission comes from a source with the size   $R_v\sim R_{jet}\ge 10^{17}$~cm  from the black hole (as in the modelling of \citet{2022MNRAS.510.2344A}),  the density of the visible / UV photons is $n_v=10^{8}\left[L_v/10^{44}\mbox{ erg/s}\right]\left[\epsilon'/5\mbox{ eV}\right]^{-1}\left[R_v/10^{17}\mbox{ cm}\right]^{-2}$~cm$^{-3}$. The mean free path of $E'=200$~GeV \gr s through such soft photon background, $
\lambda_\gamma\simeq 10^{17}\left[L_v/10^{44}\mbox{ erg/s}\right]^{-1}\left[\epsilon'/5\mbox{ eV}\right]\left[R_v/10^{17}\mbox{ cm}\right]^{2}$~cm, is comparable to the extent of the visible / ultraviolet source. Thus, the VHE \gr\ flux can be moderately affected by the pair production on the visible / ultraviolet photon backgorund generated by the jet. The optical depth for the \gr s is $\tau_{100\ GeV}=R_v/\lambda_\gamma \simeq 1 \left[L_v/10^{44}\mbox{ erg/s}\right]\left[R_v/10^{17}\mbox{ cm}\right]^{-1}$ and the VHE \gr s can escape from the source. It is possible that a more compact visible light emission from the accretion flow still gives a sub-dominant contribution to the observed source flux.  The condition $\tau_{100\ GeV}\lesssim 1$ on transparency of the source for the 100~GeV \gr s constrains parameters of such otherwise undetectable accretion flow component: $L_v\lesssim 10^{44}[R_v/10^{17}\mbox{ cm}]$~erg/s.


The size of the VHE emission region, inferred here from the microlensing detection at 100~GeV energy, is consistent with the earlier measurement in the GeV band~\citep{b0218_vovk_neronov} and suggests the extreme compactness of the gamma-ray emission region in B0218+357. At the same time, such detection implies a stationary emission source~\citep{2015NatPh..11..664N, b0218_vovk_neronov}, inconsistent with assumptions of compact blobs in the jet, moving at relativistic velocities. Altogether this points to the location of the gamma-ray emission region in the vicinity of the central supermassive black hole.

Gravitational microlensing, operating in the entire $1-100$~GeV band, may also provide a natural explanation for the extreme hardness ($\Gamma< 1.5$) of the B0218+357 flare A spectrum. The origin of such hardness, found few rare cases, is uncertain with several explanations put forward~\citep{2012A&A...541A..31N, lefa11}. However in case of B0218+357 it may be a mere consequence of the varying with energy emission region size, resulting in an energy-dependent magnification due to microlensing. This assumption can be verified if in the next occurrence both leading and trailing flares of the source would be measured enabling a direct comparison of their spectral shapes.

The estimates of the B0218+357 gamma-ray emission size are limited by the small number of the photons, detected by Fermi/LAT during the MJD~56851--56865 flaring episode. However, the hardness of the spectrum and MAGIC detection of trailing flare in the VHE band suggest during such flares the source may be readily observable with the ground-based gamma-ray telescopes, greatly exceeding Fermi/LAT in collection area. This makes B0218+357 flares an interesting target for observations with then next-generation CTA instrument.

\bibliographystyle{aa}
\bibliography{refs}

\begin{thebibliography}{39}
\expandafter\ifx\csname natexlab\endcsname\relax\def\natexlab#1{#1}\fi

\bibitem[{{Abdo} {et~al.}(2015){Abdo}, {Ackermann}, {Ajello}, {Allafort},
  {Amin}, {Baldini}, {Barbiellini}, {Bastieri}, {Bechtol}, {Bellazzini},
  {Blandford}, {Bonamente}, {Borgland}, {Bregeon}, {Brigida}, {Buehler},
  {Bulmash}, {Buson}, {Caliandro}, {Cameron}, {Caraveo}, {Cavazzuti}, {Cecchi},
  {Charles}, {Cheung}, {Chiang}, {Chiaro}, {Ciprini}, {Claus}, {Cohen-Tanugi},
  {Conrad}, {Corbet}, {Cutini}, {D'Ammando}, {de Angelis}, {de Palma},
  {Dermer}, {Drell}, {Drlica-Wagner}, {Favuzzi}, {Finke}, {Focke}, {Fukazawa},
  {Fusco}, {Gargano}, {Gasparrini}, {Gehrels}, {Giglietto}, {Giordano},
  {Giroletti}, {Glanzman}, {Grenier}, {Grove}, {Guiriec}, {Hadasch},
  {Hayashida}, {Hays}, {Hughes}, {Inoue}, {Jackson}, {Jogler},
  {J{\'o}hannesson}, {Johnson}, {Kamae}, {Kn{\"o}dlseder}, {Kuss}, {Lande},
  {Larsson}, {Latronico}, {Longo}, {Loparco}, {Lott}, {Lovellette}, {Lubrano},
  {Madejski}, {Mazziotta}, {Mehault}, {Michelson}, {Mizuno}, {Monzani},
  {Morselli}, {Moskalenko}, {Murgia}, {Nemmen}, {Nuss}, {Ohno}, {Ohsugi},
  {Paneque}, {Perkins}, {Pesce-Rollins}, {Piron}, {Pivato}, {Porter},
  {Rain{\`o}}, {Rando}, {Razzano}, {Reimer}, {Reimer}, {Reyes}, {Ritz},
  {Romoli}, {Roth}, {Saz Parkinson}, {Sgr{\`o}}, {Siskind}, {Spandre},
  {Spinelli}, {Takahashi}, {Takeuchi}, {Tanaka}, {Thayer}, {Thayer},
  {Thompson}, {Tibaldo}, {Tinivella}, {Torres}, {Tosti}, {Troja}, {Tronconi},
  {Usher}, {Vandenbroucke}, {Vasileiou}, {Vianello}, {Vitale}, {Waite},
  {Werner}, {Winer}, \& {Wood}}]{2015ApJ...799..143A}
{Abdo}, A.~A., {Ackermann}, M., {Ajello}, M., {et~al.} 2015, \apj, 799, 143

\bibitem[{Abdollahi {et~al.}(2020)Abdollahi, Acero, Ackermann, Ajello, Atwood,
  Axelsson, Baldini, Ballet, Barbiellini, Bastieri, Gonzalez, Bellazzini,
  Berretta, Bissaldi, Blandford, Bloom, Bonino, Bottacini, Brandt, Bregeon,
  Bruel, Buehler, Burnett, Buson, Cameron, Caputo, Caraveo, Casandjian, Castro,
  Cavazzuti, Charles, Chaty, Chen, Cheung, Chiaro, Ciprini, Cohen-Tanugi,
  Cominsky, Coronado-Bl{\'{a}}zquez, Costantin, Cuoco, Cutini, D'Ammando,
  DeKlotz, de~la Torre~Luque, de~Palma, Desai, Digel, Lalla, Mauro, Venere,
  Dom{\'{\i}}nguez, Dumora, Dirirsa, Fegan, Ferrara, Franckowiak, Fukazawa,
  Funk, Fusco, Gargano, Gasparrini, Giglietto, Giommi, Giordano, Giroletti,
  Glanzman, Green, Grenier, Griffin, Grondin, Grove, Guiriec, Harding, Hayashi,
  Hays, Hewitt, Horan, J{\'{o}}hannesson, Johnson, Kamae, Kerr, Kocevski,
  Kovac'evic', Kuss, Landriu, Larsson, Latronico, Lemoine-Goumard, Li,
  Liodakis, Longo, Loparco, Lott, Lovellette, Lubrano, Madejski, Maldera,
  Malyshev, Manfreda, Marchesini, Marcotulli, Mart{\'{\i}}-Devesa, Martin,
  Massaro, Mazziotta, McEnery, Mereu, Meyer, Michelson, Mirabal, Mizuno,
  Monzani, Morselli, Moskalenko, Negro, Nuss, Ojha, Omodei, Orienti, Orlando,
  Ormes, Palatiello, Paliya, Paneque, Pei, Pe{\~{n}}a-Herazo, Perkins, Persic,
  Pesce-Rollins, Petrosian, Petrov, Piron, Poon, Porter, Principe, Rain{\`{o}},
  Rando, Razzano, Razzaque, Reimer, Reimer, Remy, Reposeur, Romani, Parkinson,
  Schinzel, Serini, Sgr{\`{o}}, Siskind, Smith, Spandre, Spinelli, Strong,
  Suson, Tajima, Takahashi, Tak, Thayer, Thompson, Tibaldo, Torres, Torresi,
  Valverde, Klaveren, van Zyl, Wood, Yassine, \& Zaharijas}]{4FGL}
Abdollahi, S., Acero, F., Ackermann, M., {et~al.} 2020, The Astrophysical
  Journal Supplement Series, 247, 33

\bibitem[{{Acciari} {et~al.}(2022){Acciari}, {Ansoldi}, {Antonelli}, {Arbet
  Engels}, {Artero}, {Asano}, {Baack}, {Babi{\'c}}, {Baquero}, {Barres de
  Almeida}, {Barrio}, {Batkovi{\'c}}, {Becerra Gonz{\'a}lez}, {Bednarek},
  {Bellizzi}, {Bernardini}, {Bernardos}, {Berti}, {Besenrieder},
  {Bhattacharyya}, {Bigongiari}, {Biland}, {Blanch}, {Bonnoli},
  {Bo{\v{s}}njak}, {Busetto}, {Carosi}, {Ceribella}, {Cerruti}, {Chai},
  {Chilingarian}, {Cikota}, {Colak}, {Colombo}, {Contreras}, {Cortina},
  {Covino}, {D'Amico}, {D'Elia}, {Da Vela}, {Dazzi}, {De Angelis}, {De Lotto},
  {Delfino}, {Delgado}, {Delgado Mendez}, {Depaoli}, {Di Pierro}, {Di Venere},
  {Do Souto Espi{\~n}eira}, {Dominis Prester}, {Donini}, {Dorner}, {Doro},
  {Elsaesser}, {Fallah Ramazani}, {Fattorini}, {Ferrara}, {Fonseca}, {Font},
  {Fruck}, {Fukami}, {Garc{\'\i}a L{\'o}pez}, {Garczarczyk}, {Gasparyan},
  {Gaug}, {Giglietto}, {Giordano}, {Gliwny}, {Godinovi{\'c}}, {Green}, {Green},
  {Hadasch}, {Hahn}, {Heckmann}, {Herrera}, {Hoang}, {Hrupec}, {H{\"u}tten},
  {Inada}, {Inoue}, {Ishio}, {Iwamura}, {Jim{\'e}nez}, {Jormanainen}, {Jouvin},
  {Kajiwara}, {Karjalainen}, {Kerszberg}, {Kobayashi}, {Kubo}, {Kushida},
  {Lamastra}, {Lelas}, {Leone}, {Lindfors}, {Lombardi}, {Longo},
  {L{\'o}pez-Coto}, {L{\'o}pez-Moya}, {L{\'o}pez-Oramas}, {Loporchio}, {Machado
  de Oliveira Fraga}, {Maggio}, {Majumdar}, {Makariev}, {Mallamaci}, {Maneva},
  {Manganaro}, {Mannheim}, {Maraschi}, {Mariotti}, {Mart{\'\i}nez}, {Mazin},
  {Menchiari}, {Mender}, {Mi{\'c}anovi{\'c}}, {Miceli}, {Miener}, {Minev},
  {Miranda}, {Mirzoyan}, {Molina}, {Moralejo}, {Morcuende}, {Moreno},
  {Moretti}, {Neustroev}, {Nigro}, {Nilsson}, {Nishijima}, {Noda}, {Nozaki},
  {Ohtani}, {Oka}, {Otero-Santos}, {Paiano}, {Palatiello}, {Paneque},
  {Paoletti}, {Paredes}, {Pavleti{\'c}}, {Pe{\~n}il}, {Perennes}, {Persic},
  {Prada Moroni}, {Prandini}, {Priyadarshi}, {Puljak}, {Rhode}, {Rib{\'o}},
  {Rico}, {Righi}, {Rugliancich}, {Saha}, {Sahakyan}, {Saito}, {Sakurai},
  {Satalecka}, {Saturni}, {Schleicher}, {Schmidt}, {Schweizer}, {Sitarek},
  {{\v{S}}nidari{\'c}}, {Sobczynska}, {Spolon}, {Stamerra}, {Strom}, {Strzys},
  {Suda}, {Suri{\'c}}, {Takahashi}, {Tavecchio}, {Temnikov}, {Terzi{\'c}},
  {Teshima}, {Tosti}, {Truzzi}, {Tutone}, {Ubach}, {van Scherpenberg}, {Vanzo},
  {Vazquez Acosta}, {Ventura}, {Verguilov}, {Vigorito}, {Vitale}, {Vovk},
  {Will}, {Wunderlich}, {Zari{\'c}}, {de Palma}, {D'Ammando}, {Barnacka},
  {Sahu}, {Hodges}, {Hovatta}, {Kiehlmann}, {Max-Moerbeck}, {Readhead},
  {Reeves}, {Pearson}, {L{\"a}hteenm{\"a}ki}, {Bj{\"o}rklund}, {Tornikoski},
  {Tammi}, {Suutarinen}, {Hada}, \& {Niinuma}}]{2022MNRAS.510.2344A}
{Acciari}, V.~A., {Ansoldi}, S., {Antonelli}, L.~A., {et~al.} 2022, \mnras,
  510, 2344

\bibitem[{Aharonian {et~al.}(2007)}]{Aharonian:2007ig}
Aharonian, F. {et~al.} 2007, Astrophys. J. Lett., 664, L71

\bibitem[{{Ahnen} {et~al.}(2016){Ahnen}, {Ansoldi}, {Antonelli}, {Antoranz},
  {Arcaro}, {Babic}, {Banerjee}, {Bangale}, {Barres de Almeida}, {Barrio},
  {Becerra Gonz{\'a}lez}, {Bednarek}, {Bernardini}, {Berti}, {Biasuzzi},
  {Biland}, {Blanch}, {Bonnefoy}, {Bonnoli}, {Borracci}, {Bretz}, {Buson},
  {Carosi}, {Chatterjee}, {Clavero}, {Colin}, {Colombo}, {Contreras},
  {Cortina}, {Covino}, {Da Vela}, {Dazzi}, {De Angelis}, {De Lotto}, {de
  O{\~n}a Wilhelmi}, {Di Pierro}, {Doert}, {Dom{\'\i}nguez}, {Dominis Prester},
  {Dorner}, {Doro}, {Einecke}, {Eisenacher Glawion}, {Elsaesser},
  {Engelkemeier}, {Fallah Ramazani}, {Fern{\'a}ndez-Barral}, {Fidalgo},
  {Fonseca}, {Font}, {Frantzen}, {Fruck}, {Galindo}, {Garc{\'\i}a L{\'o}pez},
  {Garczarczyk}, {Garrido Terrats}, {Gaug}, {Giammaria}, {Godinovi{\'c}},
  {Gora}, {Guberman}, {Hadasch}, {Hahn}, {Hayashida}, {Herrera}, {Hose},
  {Hrupec}, {Hughes}, {Idec}, {Kodani}, {Konno}, {Kubo}, {Kushida}, {La
  Barbera}, {Lelas}, {Lindfors}, {Lombardi}, {Longo}, {L{\'o}pez},
  {L{\'o}pez-Coto}, {Majumdar}, {Makariev}, {Mallot}, {Maneva}, {Manganaro},
  {Mannheim}, {Maraschi}, {Marcote}, {Mariotti}, {Mart{\'\i}nez}, {Mazin},
  {Menzel}, {Miranda}, {Mirzoyan}, {Moralejo}, {Moretti}, {Nakajima},
  {Neustroev}, {Niedzwiecki}, {Nievas Rosillo}, {Nilsson}, {Nishijima}, {Noda},
  {Nogu{\'e}s}, {Paiano}, {Palacio}, {Palatiello}, {Paneque}, {Paoletti},
  {Paredes}, {Paredes-Fortuny}, {Pedaletti}, {Peresano}, {Perri}, {Persic},
  {Poutanen}, {Prada Moroni}, {Prandini}, {Puljak}, {Garcia}, {Reichardt},
  {Rhode}, {Rib{\'o}}, {Rico}, {Saito}, {Satalecka}, {Schroeder}, {Schweizer},
  {Shore}, {Sillanp{\"a}{\"a}}, {Sitarek}, {Snidaric}, {Sobczynska},
  {Stamerra}, {Strzys}, {Suri{\'c}}, {Takalo}, {Tavecchio}, {Temnikov},
  {Terzi{\'c}}, {Tescaro}, {Teshima}, {Torres}, {Toyama}, {Treves}, {Vanzo},
  {Verguilov}, {Vovk}, {Ward}, {Will}, {Wu}, {Zanin}, \&
  {Desiante}}]{2016A&A...595A..98A}
{Ahnen}, M.~L., {Ansoldi}, S., {Antonelli}, L.~A., {et~al.} 2016, \aap, 595,
  A98

\bibitem[{{Aleksi{\'c}} {et~al.}(2014){Aleksi{\'c}}, {Ansoldi}, {Antonelli},
  {Antoranz}, {Babic}, {Bangale}, {Barrio}, {Gonz{\'a}lez}, {Bednarek},
  {Bernardini}, {Biasuzzi}, {Biland}, {Blanch}, {Bonnefoy}, {Bonnoli},
  {Borracci}, {Bretz}, {Carmona}, {Carosi}, {Colin}, {Colombo}, {Contreras},
  {Cortina}, {Covino}, {Da Vela}, {Dazzi}, {De Angelis}, {De Caneva}, {De
  Lotto}, {Wilhelmi}, {Mendez}, {Prester}, {Dorner}, {Doro}, {Einecke},
  {Eisenacher}, {Elsaesser}, {Fonseca}, {Font}, {Frantzen}, {Fruck}, {Galindo},
  {L{\'o}pez}, {Garczarczyk}, {Terrats}, {Gaug}, {Godinovi{\'c}}, {Mu{\~n}oz},
  {Gozzini}, {Hadasch}, {Hanabata}, {Hayashida}, {Herrera}, {Hildebrand},
  {Hose}, {Hrupec}, {Idec}, {Kadenius}, {Kellermann}, {Kodani}, {Konno},
  {Krause}, {Kubo}, {Kushida}, {La Barbera}, {Lelas}, {Lewandowska},
  {Lindfors}, {Lombardi}, {Longo}, {L{\'o}pez}, {L{\'o}pez-Coto},
  {L{\'o}pez-Oramas}, {Lorenz}, {Lozano}, {Makariev}, {Mallot}, {Maneva},
  {Mankuzhiyil}, {Mannheim}, {Maraschi}, {Marcote}, {Mariotti},
  {Mart{\'\i}nez}, {Mazin}, {Menzel}, {Miranda}, {Mirzoyan}, {Moralejo},
  {Munar-Adrover}, {Nakajima}, {Niedzwiecki}, {Nilsson}, {Nishijima}, {Noda},
  {Orito}, {Overkemping}, {Paiano}, {Palatiello}, {Paneque}, {Paoletti},
  {Paredes}, {Paredes-Fortuny}, {Persic}, {Poutanen}, {Moroni}, {Prandini},
  {Puljak}, {Reinthal}, {Rhode}, {Rib{\'o}}, {Rico}, {Garcia}, {R{\"u}gamer},
  {Saito}, {Saito}, {Satalecka}, {Scalzotto}, {Scapin}, {Schultz}, {Schweizer},
  {Shore}, {Sillanp{\"a}{\"a}}, {Sitarek}, {Snidaric}, {Sobczynska}, {Spanier},
  {Stamatescu}, {Stamerra}, {Steinbring}, {Storz}, {Strzys}, {Takalo},
  {Takami}, {Tavecchio}, {Temnikov}, {Terzi{\'c}}, {Tescaro}, {Teshima},
  {Thaele}, {Tibolla}, {Torres}, {Toyama}, {Treves}, {Uellenbeck}, {Vogler},
  {Zanin}, {Kadler}, {Schulz}, {Ros}, {Bach}, {Krau{\ss}}, \&
  {Wilms}}]{2014Sci...346.1080A}
{Aleksi{\'c}}, J., {Ansoldi}, S., {Antonelli}, L.~A., {et~al.} 2014, Science,
  346, 1080

\bibitem[{{Atwood} {et~al.}(2009){Atwood}, {Abdo}, {Ackermann}, \&
  et~al.}]{Fermi_LAT}
{Atwood}, W.~B., {Abdo}, A.~A., {Ackermann}, M., \& et~al. 2009, \apj, 697,
  1071

\bibitem[{Berge {et~al.}(2006)}]{Berge:2006uls}
Berge, D. {et~al.} 2006, Science, 314, 1424

\bibitem[{{Biggs} \& {Browne}(2018)}]{2018MNRAS.476.5393B}
{Biggs}, A.~D. \& {Browne}, I.~W.~A. 2018, \mnras, 476, 5393

\bibitem[{{Blackburne} {et~al.}(2011){Blackburne}, {Pooley}, {Rappaport}, \&
  {Schechter}}]{2011ApJ...729...34B}
{Blackburne}, J.~A., {Pooley}, D., {Rappaport}, S., \& {Schechter}, P.~L. 2011,
  \apj, 729, 34

\bibitem[{{Buson} {et~al.}(2015){Buson}, {Cheung}, {Larsson}, \&
  {Scargle}}]{B0218+35_Fermi_paper_new}
{Buson}, S., {Cheung}, C.~C., {Larsson}, S., \& {Scargle}, J.~D. 2015, arXiv
  e-prints, arXiv:1502.03134

\bibitem[{{Cheung} {et~al.}(2014){Cheung}, {Larsson}, {Scargle}, {Amin},
  {Blandford}, {Bulmash}, {Chiang}, {Ciprini}, {Corbet}, {Falco}, {Marshall},
  {Wood}, {Ajello}, {Bastieri}, {Chekhtman}, {D'Ammando}, {Giroletti}, {Grove},
  {Lott}, {Ojha}, {Orienti}, {Perkins}, {Razzano}, {Smith}, {Thompson}, \&
  {Wood}}]{2014ApJ...782L..14C}
{Cheung}, C.~C., {Larsson}, S., {Scargle}, J.~D., {et~al.} 2014, \apjl, 782,
  L14

\bibitem[{{Cohen} {et~al.}(2000){Cohen}, {Hewitt}, {Moore}, \&
  {Haarsma}}]{2000ApJ...545..578C}
{Cohen}, A.~S., {Hewitt}, J.~N., {Moore}, C.~B., \& {Haarsma}, D.~B. 2000,
  \apj, 545, 578

\bibitem[{{Courvoisier}(1998)}]{1998A&ARv...9....1C}
{Courvoisier}, T. J.~L. 1998, \aapr, 9, 1

\bibitem[{{Dai} {et~al.}(2010){Dai}, {Kochanek}, {Chartas}, {Koz{\l}owski},
  {Morgan}, {Garmire}, \& {Agol}}]{2010ApJ...709..278D}
{Dai}, X., {Kochanek}, C.~S., {Chartas}, G., {et~al.} 2010, \apj, 709, 278

\bibitem[{{Ghisellini} \& {Tavecchio}(2009)}]{2009MNRAS.397..985G}
{Ghisellini}, G. \& {Tavecchio}, F. 2009, \mnras, 397, 985

\bibitem[{{Giannios} {et~al.}(2009){Giannios}, {Uzdensky}, \&
  {Begelman}}]{2009MNRAS.395L..29G}
{Giannios}, D., {Uzdensky}, D.~A., \& {Begelman}, M.~C. 2009, \mnras, 395, L29

\bibitem[{Gould \& Schr\'eder(1967)}]{PhysRev.155.1404}
Gould, R.~J. \& Schr\'eder, G.~P. 1967, Phys. Rev., 155, 1404

\bibitem[{{Lefa} {et~al.}(2011){Lefa}, {Rieger}, \& {Aharonian}}]{lefa11}
{Lefa}, E., {Rieger}, F.~M., \& {Aharonian}, F. 2011, \apj, 740, 64

\bibitem[{{Levinson}(2000)}]{2000PhRvL..85..912L}
{Levinson}, A. 2000, \prl, 85, 912

\bibitem[{{Mittal} {et~al.}(2007){Mittal}, {Porcas}, \&
  {Wucknitz}}]{2007A&A...465..405M}
{Mittal}, R., {Porcas}, R., \& {Wucknitz}, O. 2007, \aap, 465, 405

\bibitem[{{Narayan} \& {Yi}(1994)}]{1994ApJ...428L..13N}
{Narayan}, R. \& {Yi}, I. 1994, \apjl, 428, L13

\bibitem[{Neronov(2019)}]{Neronov:2019uht}
Neronov, A. 2019, J. Phys. Conf. Ser., 1263, 012001

\bibitem[{{Neronov} \& {Aharonian}(2007)}]{2007ApJ...671...85N}
{Neronov}, A. \& {Aharonian}, F.~A. 2007, \apj, 671, 85

\bibitem[{{Neronov} \& {Semikoz}(2003)}]{2003NewAR..47..693N}
{Neronov}, A. \& {Semikoz}, D. 2003, \nar, 47, 693

\bibitem[{Neronov {et~al.}(2008)Neronov, Semikoz, \&
  Sibiryakov}]{Neronov:2008ih}
Neronov, A., Semikoz, D., \& Sibiryakov, S. 2008, Mon. Not. Roy. Astron. Soc.,
  391, 949

\bibitem[{{Neronov} {et~al.}(2012){Neronov}, {Semikoz}, \&
  {Taylor}}]{2012A&A...541A..31N}
{Neronov}, A., {Semikoz}, D., \& {Taylor}, A.~M. 2012, \aap, 541, A31

\bibitem[{{Neronov} {et~al.}(2015){Neronov}, {Vovk}, \&
  {Malyshev}}]{2015NatPh..11..664N}
{Neronov}, A., {Vovk}, I., \& {Malyshev}, D. 2015, Nature Physics, 11, 664

\bibitem[{{Patnaik} {et~al.}(1995){Patnaik}, {Porcas}, \&
  {Browne}}]{1995MNRAS.274L...5P}
{Patnaik}, A.~R., {Porcas}, R.~W., \& {Browne}, I.~W.~A. 1995, \mnras, 274, L5

\bibitem[{Peterson(2006)}]{blr}
Peterson, B. 2006, Lecture Notes in Physics, 693

\bibitem[{{Poutanen} \& {Stern}(2010)}]{2010ApJ...717L.118P}
{Poutanen}, J. \& {Stern}, B. 2010, \apjl, 717, L118

\bibitem[{{Ptitsyna} \& {Neronov}(2016)}]{2016A&A...593A...8P}
{Ptitsyna}, K. \& {Neronov}, A. 2016, \aap, 593, A8

\bibitem[{{Rees} {et~al.}(1982){Rees}, {Begelman}, {Blandford}, \&
  {Phinney}}]{1982Natur.295...17R}
{Rees}, M.~J., {Begelman}, M.~C., {Blandford}, R.~D., \& {Phinney}, E.~S. 1982,
  \nat, 295, 17

\bibitem[{{Shakura} \& {Sunyaev}(1973)}]{1973A&A....24..337S}
{Shakura}, N.~I. \& {Sunyaev}, R.~A. 1973, \aap, 24, 337

\bibitem[{{Vovk} \& {Neronov}(2013)}]{2013ApJ...767..103V}
{Vovk}, I. \& {Neronov}, A. 2013, \apj, 767, 103

\bibitem[{{Vovk} \& {Neronov}(2016)}]{b0218_vovk_neronov}
{Vovk}, I. \& {Neronov}, A. 2016, \aap, 586, A150

\bibitem[{{Wilkes}(2004)}]{2004ASPC..311...37W}
{Wilkes}, B. 2004, in Astronomical Society of the Pacific Conference Series,
  Vol. 311, AGN Physics with the Sloan Digital Sky Survey, ed. G.~T. {Richards}
  \& P.~B. {Hall}, 37

\bibitem[{{Wood} {et~al.}(2017){Wood}, {Caputo}, {Charles}, {Di Mauro},
  {Magill}, {Perkins}, \& {Fermi-LAT Collaboration}}]{fermipy}
{Wood}, M., {Caputo}, R., {Charles}, E., {et~al.} 2017, in International Cosmic
  Ray Conference, Vol. 301, 35th International Cosmic Ray Conference
  (ICRC2017), 824

\bibitem[{{Yuan} \& {Narayan}(2014)}]{2014ARA&A..52..529Y}
{Yuan}, F. \& {Narayan}, R. 2014, \araa, 52, 529

\end{thebibliography}


\end{document}